\documentstyle[12pt]{article}

\def\Journal#1#2#3#4{{#1} {\bf #2}, #3 (#4)}

\def\NPB{{\em Nucl. Phys.} B}
\def\PLB{{\em Phys. Lett.}  B}
\def\PRL{\em Phys. Rev. Lett.}
\def\PRD{{\em Phys. Rev.} D}

\def\be{\begin{equation}}
\def\ee{\end{equation}}
\def\bea{\begin{eqnarray}}
\def\eea{\end{eqnarray}}

\normalsize
\begin{document}

\title{THE ROME APPROACH TO CHIRALITY}

\author{ M. TESTA\\
Dip. di Fisica, Universit\'{a} di Roma
``La Sapienza``\\ and\\ INFN Sezione di Roma I\\
Piazzale Aldo Moro 2\\I-00185 Roma\\
 ITALY\\testa@roma1.infn.it\\}
\maketitle
\begin{abstract}
Some general considerations
on the problem of non perturbative definition of
Chiral Gauge Theories are presented and exemplified
within the particular proposal known as the  Rome
Approach.
\end{abstract}
  
\section{Introduction}\label{sec:intr}

Gauge invariance starts as a classical concept:
Vector and Chiral symmetries are on the same ground.
In Quantum Field Theory, on the contrary, there is
a deep difference between them,  due to the lack of
a chiral invariant regularization.

This fact is not merely a mathematical fancy, but
is subject to direct  experimental observation,
e.g. in the $\pi_0\rightarrow \gamma \gamma$ decay
and similar. Also, the global structure of the
Standard Model is deeply affected by the non 
existence of a gauge invariant regularization. In
fact already in QCD, although a Vector Theory, the
Chiral Classification  of  Local Observables
(Current Algebra) is a complicated problem. Non
perturbative computations in Chiral Gauge Theories
could clarify fundamental issues, as  the
possibility of dynamical Higgs mechanism, Baryon
non conservation, the question of  Naturalness.

	How can we quantize chiral gauge theories?

Several approaches have been explored:

	1) Non gauge invariant
quantization{\cite{borrelli1} \cite{borrelli2}}
(Rome approach\footnote{Within this class falls
also the formulation of the Zaragoza 
group{\cite{alonso}}.}) based on the Bogolubov
method

	2) Gauge invariant quantization

	    (J.Smit and P.Swift\cite{smit},
S.Aoki\cite{aoki},.....)

	3) Other degrees of freedom....

	    Mirror Fermions (I.Montvay\cite{montvay})

	4)....and other dimensions

	   (D.Kaplan\cite{kaplan}, R.Narayanan and
H.Neuberger\cite{nn},  S.Randjbar-Daemi and
J.Strathdee\cite{rds})

	5) Fine-Grained Fermions

	    (G.Schierholz\cite{sch}, G.Ôt Hooft\cite{hoo}, 
P.Hernandez and R.Sundrum\cite{hs})

\section{{Quantization of Chiral Gauge
Theories}}\label{sec:quantization}

In order to quantize a theory we have to go through
several steps:

$\bullet$ Definition of the Target Theory

$\bullet$ Regularization

$\bullet$ Renormalization

\subsection{Target Theory}

We have, first of all, to decide what is the theory
we are aiming at, i.e. the so called Target Theory.
The formal (continuum) theory we want to reproduce
is:

\begin{eqnarray} {\it L}={\it L}_{G}+{\it
L}_{gf}\nonumber\\ {\it
L}_{G}=\bar{\psi}_{L}\widehat{D}\psi_{L}
+\frac{1}{4}F_{\mu\nu}^{a}F_{\mu\nu}^{a}+
\bar{\psi}_{R}\widehat{\partial}\psi_{R}
\label{target}\\ {\it
L}_{gf}=\frac{1}{2\alpha_{0}}(\partial_{\mu}A_{\mu}^{a})
(\partial_{\nu}A_{\nu}^{a})+\bar
c\partial_{\mu}D_{\mu}c\nonumber
\end{eqnarray} where $D_{\mu}$ denotes the
covariant derivative
\begin{eqnarray}
D_{\mu}=\partial_{\mu}+ig_{0}A_{\mu}^{a}T^{a}
\label{covariant}
\end{eqnarray}

In Eq.(\ref{covariant}), the $T^{a}$'s are the
appropriate generators of the  gauge group $G$,
$g_{0}$ and $\alpha_{0}$ denote the bare coupling
and  gauge fixing parameter, respectively. The rest
of the notation is self-explanatory.

A few comments are in order here:

$\bullet$	Presence of Gauge Fixing

As we will see later, it is rather difficult to
dispose of it.  Our general attitude is that it
makes no harm. We are aware of a general argument
 by Neuberger\cite{neubzero} which shows that the
sum over Gribov copies on a finite lattice  is such
that the expectation value of any gauge invariant
quantity  assumes the form of an indeterminate
expression $0\over 0$. Neuberger argument applies,
however, in situations in which the lattice
regularization is {\it exactly} gauge  invariant.
In the present case gauge invariance is recovered
only in continuum limit and a crucial  ingredient
of the argument, i.e. compactness, is lost.  Of
course this point deserves further investigation.

$\bullet$	No Higgs degrees of freedom are present
in Eq.(\ref{target}),  but they could be easily
added.

$\bullet$ The particular gauge group $SU(2)$ has
been considered in order  to avoid Local Anomalies
without the need to introduce other fermions. Of
course such a theory  is probably affected by the
Witten Global Anomaly\cite{witten}, but this, of
course,  does not show up in perturbative checks of
the method.

$\bullet$ The presence of fictitious, non
interacting degrees of freedom, $\psi_R$,  is
useful to limit the form of the counterterms. They
complicate the issue of  Dynamical Fermion 
Non-Conservation and will be disposed off later.

The most important informations encoded in the
Target Theory, are represented by its symmetries.
In the present case they are:

a)	BRST\cite{brst}

If we write the gauge fixing in the linearized form:
\be {\it
L}_{gf}=\frac{\alpha_{0}}{2}(\lambda^{a}\lambda^{a})
+i\lambda^{a} (\partial_{\mu}A_{\mu}^{a})+\bar{c}
\partial_{\mu}D_{\mu}c
\label{linearizzata}\ee it can be shown that ${\it
L}_{G}$ and ${\it L}_{gf}$ are  separately invariant
under the BRST transformations, defined on the
basic fields as
\begin{eqnarray}
\delta\psi_{L}\equiv\epsilon\delta_{BRST}\psi_{L}=i\epsilon 
g_{0}c^{a}T^{a} \psi_{L}\nonumber\\
\delta\bar{\psi}_{L}\equiv\epsilon\delta_{BRST}\bar{\psi}_{L}=i\epsilon 
g_{0}\bar{ \psi}_{L} T^{a}c^{a} \nonumber\\
\delta\psi_{R}=\delta\bar{\psi}_{R}=0 \nonumber\\
\delta
A_{\mu}^{a}\equiv\epsilon\delta_{BRST}A_{\mu}^{a}=\epsilon
(D_{\mu}c)^{a} \nonumber\\
\delta\lambda^{a}=0
\label{BRST}\\
\delta c^{a}\equiv
\epsilon\delta_{BRST}c^{a}=-\frac{1}{2}\epsilon
g_{0} f_{abc}c^{b}c^{c}\nonumber\\
\delta \bar{c}^{a}\equiv
\epsilon\delta_{BRST}\bar{c}^{a}=
\epsilon\lambda^{a}\nonumber
\end{eqnarray} where $\epsilon$ is a grassmannian
parameter. In fact ${\it L}_{gf}$ is automatically
BRST invariant as a consequence of nilpotency:
\be
\delta_{BRST}^2=0
\ee

Other (global) Symmetries.

b) Vector-like:
\begin{eqnarray}
\psi _L\to V\psi _L\nonumber\\
\psi _R\to V\psi _R\nonumber\\ A_\mu \to VA_\mu
V^+\label{global}\\ V\in G\nonumber
\end{eqnarray}

c) Shift Symmetry, that is the symmetry under the
shift of the antighost field:
\begin{eqnarray}
\bar{c}(x)\rightarrow \bar{c}(x)+const.
\label{shift}\end{eqnarray}

d) Global rotation of the right handed fields:
\bea
\psi _R\to V\psi _R\nonumber\\
\psi _L\to \psi _L\label{obs}\\ A_\mu \to
A_\mu\nonumber
\eea As usual, the invariance of the lagrangian
implies an infinite set of identities on the
Green's Functions:
\be
\left\langle {\Phi _1(x_1)\ldots \ldots \Phi
_n(x_n)}
 \right\rangle \equiv \int {d\mu \;e^{S_{cl}}\Phi
_1(x_1)\ldots \ldots \Phi _n(x_n)\;}\label{green}
\ee In particular BRST invariance implies:
\be
\left\langle {\delta _{BRST}\left( {\Phi
_1(x_1)\ldots \ldots \Phi _n(x_n)} \right)}
\right\rangle
 =0\label{brstinvariance}
\ee

\subsection{Regularization}

Once the Target Theory has been defined, in order
to set up a consistent quantizaton scheme,  a
regulator must be introduced. All the following
considerations are not tied to
 a particular regularization. They are quite
general  features of any known regularization
scheme. However lattice discretization is very
peculiar since it also allows the rather unique
opportunity to perform systematic
 nonperturbative numerical explorations. This is
why, in the following, we will
 exemplify the Rome approach in a Lattice
Discretization setup.

Therefore, we first of all regularize the theory
discretizing it in presence of gauge fixing:
\begin{eqnarray} {\it L}_{0} =  (\frac{1}{2a})
\sum_{\mu}[\bar{\psi}_{L}(x+\mu)U_{\mu}(x)\gamma_{\mu}\psi_{L}(x)\nonumber\\
+\bar{\psi}_{R}(x+\mu)\gamma_{\mu}\psi_{R}(x) +
h.c.]
\label{naive}\\
+(\frac{1}{2a^4g_{0}^2}\sum_{\mu,\nu}Tr(P_{\mu,\nu}-1)
\nonumber
\end{eqnarray} where $P_{\mu,\nu}$ denotes the
plaquette formed with the link variables 
$U_{\mu}$.

The general difficulties inherent to the
quantization of a Chiral Gauge Theory,  manifest
themselves, in this case, in the form of the so
called Doubling Problem.

In fact the naive discretization of the Dirac
action in Eq.(\ref{naive}) leads to a (inverse)
 Fermion Propagator of the form:

\begin{eqnarray} S^{-1}(p)=\frac{1}{a}
\sum_{\mu,=1}^{4}\gamma_{\mu}\sin{(ap_\mu)}\label{doubling}
\end{eqnarray}

The problem with Eq.(\ref{doubling}) is that it
implies an unwanted
 proliferation of Fermion species usually referred
to as the Doubling Problem. A general solution has
been proposed by Wilson\cite{wilson} and it
consists in adding to the fermion action the
so-called Wilson term:
\begin{eqnarray} {\it L}_{W} =
(\frac{-r}{2a})\sum_{\mu}\{
[\bar{\psi}_{L}(x+\mu)\psi_{R}(x)\nonumber\\
+\bar{\psi}_{L}(x)\psi_{R}(x+\mu) -
2\bar{\psi}_{L}(x)\psi_{R}(x)] + h.c.\}
\label{Wilson}\end{eqnarray} which reads, in the
continuum, as:
\begin{eqnarray}
\it{L_W} \approx a r\bar{\psi}_L\Box\psi_R + h.c.
\end{eqnarray} The presence of the Wilson term
modifies Eq.(\ref{doubling}) as:
\begin{eqnarray} S^{-1}(p)=\frac{1}{a}
\sum_{\mu,=1}^{4}\gamma_{\mu}\sin{(ap_\mu)}+\frac{r}{a}\sum_{\mu,=1}^{4}\sin^2{(\frac{ap_\mu}{2})}
\end{eqnarray} In this way the doubling problem is
avoided.

However the Wilson term leaves us with an unwanted
chiral symmetry breaking. This is a very general
fact as expressed by the:

{\bf Nielsen-Ninomiya Theorem\cite{nielseninomiya}:

Any local, chiral symmetric, bilinear action
implies Spectrum Doubling.}

The whole problem of quantizing Chiral Gauge
Theories is precisely to understand the effect of
such regularization-induced chirality breaking.

\subsection{Regularization-Induced Symmetry
Breaking}

In the language of renormalization theory, ${\it
L_W}$ is a so-called ``irrelevant`` term:  its
presence can be compensated by finite or power
divergent counterterms. 

Since this a central point (and a very inconvenient
one) let us discuss in more detail  the origin and
the manifestation of this phenomenon.

We start with a theory whith a given symmetry group
$G$, broken at the level of the cutoff,  say $a$.

As an example we may think of a $\lambda{\phi}^4$
theory (symmetric under $\phi\rightarrow -\phi$) 
with an additional $O_5\approx{\phi}^5$ term in the
lagrangian.

Of course, in order to be really a correction of
order $a$ to start with, $O_5$ should be a finite 
(i.e. renormalized) operator in the continuum limit
($a\rightarrow 0$) in order to avoid  an immediate
back-reaction giving rise to counter-terms $\phi$
and ${\phi}^3$.

The theory is defined by the action:
\be S(\phi )=S_{Sym}(\phi )+a\int {d^4xO_5(x)}
\ee

We can now expand any Green's function in powers of
$a \int{dx O_5}$ and consider,  as a particular
example,  the three-point Green's function
(expected to vanish in the symmetric theory):
\begin{eqnarray}
\langle \phi(x_1) \phi(x_2) \phi(x_3) \rangle
=\nonumber\\
\sum_{n=0}^{\infty}\frac{1}{n!}\langle \phi(x_1)
\phi(x_2) \phi(x_3) (a \int{dx O_5})^n \rangle
\end{eqnarray}

First order correction:
\begin{eqnarray}
\langle {\phi(x_1) \phi(x_2) \phi(x_3)}
{\rangle}_{(1)} = a \int dy\langle \phi(x_1)
\phi(x_2) \phi(x_3) O_5(y) \rangle
\end{eqnarray}

This is fine (i.e. ${\rightarrow}_{a\to 0} 0$)
since we assumed that the single  insertion of
$O_5$ is finite.

The next interesting contribution, in this
particular example, is the third order correction:
\bea
\langle {\phi(x_1) \phi(x_2) \phi(x_3)}
{\rangle}_{(3)} \approx\nonumber\\
\approx a^3 \int dy_1 dy_2 dy_3 \langle \phi(x_1)
\phi(x_2) \phi(x_3) O_5(y_1) O_5(y_2)
O_5(y_3)\rangle\label{cubic}
\eea

The contribution coming from Eq.(\ref{cubic}) is
hardly of order $a^3$. In fact the integrals in the
r.h.s. of Eq.(\ref{cubic}) get contributions from
the region where the $y$'s are close together, 
which can be estimated through the Operator Product
Expansion as:
\begin{eqnarray}
\int dy_1 dy_2 dy_3 O_5(y_1) O_5(y_2)
O_5(y_3)\approx \frac{1}{a^4}\int dy_1 O_3(y_1)
\end{eqnarray} where $O_3$ is, in the present case
a renormalized version of $\phi^3$.

This integration region gives therefore rise to a
linearly divergent contribution  (as expected from
power-counting) of the form:
\begin{equation}
\langle {\phi(x_1) \phi(x_2) \phi(x_3)}
{\rangle}_{(3)}\approx 
\frac{1}{a} \int dy_1 \langle {\phi(x_1) \phi(x_2)
\phi(x_3) O_3(y_1)} {\rangle}
\end{equation}

Depending on the particular regularization
employed, the appearance of these contributions 
can be shifted to higher orders in Perturbation
Theory, but can hardly be eliminated,  unless some
{\em exact} selection rule is operating at the
level of the regularized theory.

In this situation the only way to get a sensible
continuum limit is to add to the action  all
possible symmetry breaking (and non-Lorentz
invariant, in  the case of the Lattice
Discretization) counterterms with dimension $D\le
4$.

This discussion is directly relevant to the case in
which Gauge Symmetry is violated by the regulator,
at least in the case in which the theory is defined
within some specific gauge.

\subsection{The Rome Approach to Chirality}

In the case of a Target Theory as in
Eq.(\ref{target}) we have plenty of possible
counterterms. In fact, defining the vector field
$A_{\mu}$ as e.g.:
\be ag_0A_\mu \equiv {{U_\mu -U_\mu ^+} \over {2i}}
\ee we have:

$\bullet$ Counterterms with $D=2$:
\be -{{\delta \mu _A^2} \over 2}A_\mu ^a(x)A_\mu
^a(x)
\ee No ghost mass counterterms arise because of the
shift symmetry.

$\bullet$ Counterterms with $D=3$:
\be
\delta M\left[ {\bar \psi _L(x)\psi _R(x)+h.c.}
\right]
\ee

$\bullet$ Examples of counterterms with $D=4$:

Fermion vertices counterterms:
\bea -i\delta g_R\bar \psi _RT^a A_\mu ^a\gamma
_\mu \psi _R\nonumber\\ -i\delta g_L\bar \psi _LT^a
A_\mu ^a\gamma _\mu \psi _L
\eea

Non minimal terms in $A_{\mu}$, $\bar c^a$ and
${c^c}$:
\bea
\left( {\partial _\mu A_\mu ^a} \right)^2\nonumber\\
f^{abc}\partial _\mu A_\nu ^aA_\mu ^bA_\nu ^c\\
\sum\limits_\mu  {\partial _\mu A_\mu ^a\partial
_\mu A_\mu ^a}\nonumber
\eea

\be
\delta g_{gh}f_{abc}\bar c^a\partial _\mu \left(
{A_\mu ^bc^c} \right)\label{ghost}
\ee

The presence of the counterterm Eq.(\ref{ghost}) is
very important  because it signals an irreversible
breaking  of geometry. We will come back later on
this point.

The strategy is to fix the values of the
counterterms as follows.

First of all compute (e.g. non perturbatively):
\bea
\left\langle \Phi _1(x_1)\ldots \ldots \Phi _n(x_n)
\right\rangle =\nonumber\\
 \int {DU_\mu D\bar \psi D\psi D\bar cDc\;}
  &e^{S_0+S_W-{1 \over {2\alpha _0}}\int
{d^4x\left( {\partial _\mu A_\mu ^a}
\right)^2+S_{ghost}+S_{c.t.}}}\\ 
\Phi _1(x_1)\ldots \ldots \Phi _n(x_n)\nonumber
\eea then tune the values of the counterterms
imposing the BRST ÒIdentitiesÓ
Eq.(\ref{brstinvariance}). This is at best possible
up to order $a$ (and impossible if there are
unmatched anomalies).

By this procedure we define a ÒbareÓ chiral theory
with parameters $g_0$
 and $\alpha _0$ implicitly defined by the BRST
transformations.

It is now possible to carry out the usual
non-perturbative renormalization,  by fixing the
bare parameters to reproduce given finiteness
conditions  (on physical quantities and/or Green
functions).

The procedure just outlined is completely non
perturbative. However perturbation theory may be
practically helpful. In fact the theory so defined
should be asymptotycally free and we may
distinguish  two different kinds of counterterms:

$\bullet$	Dimensionless counterterms:
\be
\delta Z=f(g_0,\alpha _0)
\ee

The value of these counterterms can be reliably
esimated from perturbation theory.

$\bullet$	Dimensionful counterterms:
\be
\delta M={1 \over a}f(g_0,\alpha _0)
\ee

These counterterms are essentially non perturbative.
In fact exponentially small contributions to $f$
can be rescued in the continuum limit by the 
factor ${1 \over a}$:
\be {1 \over a}f(g_0,\alpha _0)\approx {{e^{-\,{1
\over {g_0^2}}}} \over a}\approx \Lambda
\ee where $\Lambda$ is the usual scale defined
through dimensional transmutation.

The scheme just described has been checked in
perturbation theory at 1-loop\cite{borrelli1}  and
reproduces the results of continuum perturbation
theory.

\subsection{Are Ghosts (and Gauge-Fixing)
Unavoidable?}

G.C.Rossi, R.Sarno and R.Sisto\cite{rossi} computed
the ghost counterterm defined in Eq.(\ref{ghost})
at two loops (in dimensional regularization) and
found that $\delta g{}_{gh}\ne 0$. Thus (at least
if we trust Perturbation Theory) we cannot invert
the Faddeev-Popov  procedure and remove the gauge
fixing. This is the signal that the gauge geometry
is irreversibly lost.

\subsection{Possible Obstructions}

Several points may go wrong during the
implementation of the program just described. In
particular the procedure will not work in presence
of:

$\bullet$	Non cancelled perturbative anomalies.

$\bullet$ Non perturbative anomalies\cite{witten}.

$\bullet$ The symmetry defined by Eq.(\ref{obs})
must be realized $\grave {a}$ la Wigner. This is
not a trivial requirement as the example of QCD
clearly shows. In fact in QCD Chiral symmetry can
be recovered by an appropriate tuning of the quark
masses, but the phase in which it is recovered is a
completely dynamical issue.

\section{Gauge Averaging}

	An interesting proposal to deal with a gauge
non-invariant regulator  is the so-called method of
Gauge Averaging (D.Foerster, H.Nielsen and
M.Ninomiya\cite{foerster},
 J.Smit and P.Swift\cite{smit},
S.Aoki\cite{aoki},.....).

The basic idea is to make the Wilson term, or any
other gauge non-invariant term,  invariant through
the introduction of an additional degree of freedom
in the form of  the angular part of a scalar
Higgs-like field $\Omega (x)$ with $\Omega (x)\in
G$ as:
\be a\bar \psi _R\partial ^2\left( {\Omega ^+\psi
_L} \right)
\ee

This theory is now exactly invariant under the
gauge transformations $g\in G_1$:
\bea
\Omega \to g\,\Omega \equiv \Omega ^g\nonumber\\
\psi _L\to g\;\psi _L\equiv \psi
_L^g\label{unphysical}\\
\psi _R\to \psi _R\nonumber\\ U_\mu \to g^+(x+\mu
)\,U_\mu \,g(x)\equiv U^g_\mu\nonumber
\eea

In this way any action can be made invariant under
$G_1$:
\be
\int {DUD\psi D\bar \psi D\Omega e^{S_{NI}(U^\Omega
,\psi ^\Omega ,\bar \psi ^\Omega )}\;}
\ee

However the group $G_1$ is not the physical gauge
group because the Target Theory  does not contain
any scalar field and $\Omega (x)$ cannot be
identified with a physical  Higgs field:
 switching off the gauge interaction we should get
back a free fermion theory. Moreover the gauge
average seems to produce theories with too many
relevant parameters. We must remember, at this
point, that the correct theory should, instead,  be
invariant under the {\em physical} gauge group:
\bea
\Omega \to \,\Omega\nonumber \\
\psi _L\to g\;\psi _L\equiv \psi _L^g\\
\psi _R\to \psi _R\nonumber\\ U_\mu \to g^+(x+\mu
)\,U_\mu \,g(x)\equiv U^g_\mu\nonumber
\eea or:
\bea
\Omega \to g\Omega \equiv \Omega ^g\nonumber\\
\psi _L\to \psi _L\\
\psi _R\to \psi _R\nonumber\\ U_\mu \to U_\mu
\nonumber
\eea

If this is the case, then $\Omega(x)$ can be
transformed into the identity and  decoupled
completely.

A possible strategy\cite{onlattice} to decouple
$\Omega(x)$ is to add to the  action (and adjust) 
all the relevant $G_1$-invariant counterterms.

Among these we have, for example:
\be
\delta S\approx {\kappa  \over 2}\int {d^4x\left(
{D(A)_\mu \Omega (x)} \right)^2}
\ee which provides both a mass term for the gauge
field $A_{\mu}$ and a kinetic term for $\Omega (x)$.

It is possible to show\cite{onlattice} that the
decoupling of $\Omega (x)$ can be achieved by, 
first of all, gauge fixing the theory:
\be
\left\langle O \right\rangle ={1 \over Z}\int
{D\Omega }\int {DUD\psi D\bar \psi }D\bar cDc\\
e^{S_{NI}(U^\Omega ,\psi ^\Omega ,\bar \psi ^\Omega
)+S_{gf}(\bar c,c,U)}O\left( {U,\psi ,\bar \psi }
\right)
\ee and then tuning the parameters in such a way
that the integrand becomes 
$\Omega$-independent. This procedure turns out to
be exactly equivalent to impose  the BRST
identities.  The $\Omega$-integration can be
dropped and we are back to the Rome approach.

Suppose, instead, we try to integrate the theory
without any gauge-fixing. Then we could try to
argue as follows.

We start by decomposing the action as:
\be S=S_{GI}+a\int {d^4yW(y)}\label{effective}
\ee where $S_{GI}$ is the gauge-invariant part (the
theory with the doublers  in the physical spectrum)
and $W(x)$ is the ``irrelevant`` dimension 5 Wilson
term. If we denote by $O_{GI}$ any (multi-) local
gauge-invariant operator, we have, for its
expectation value (at least formally):
\bea
\left\langle {O_{GI}} \right\rangle \equiv \int
{DUD\psi D\bar \psi e^{S_{GI}+a\int
{dyW(y)}}O_{GI}}\equiv\nonumber\\
\equiv\sum\limits_{n=0}^\infty  {\left\langle
{O_{GI}} \right\rangle _{(n)}}
=\sum\limits_{n=0}^\infty  {{1 \over
{n!}}\left\langle {O_{GI}(a\int {d^4yW(y)})^n}
\right\rangle }=\nonumber\\
=\sum\limits_{n=0}^\infty  {{{a^n} \over {n!}}\int
{D\Omega }\left\langle {O_{GI} (\int {d^4y W^\Omega
(y)})^n} \right\rangle }\label{expansion}
\eea where we have introduced an (harmless)
integration over the fictitious variable
$\Omega(x)$. In fact this operation is well defined
within any Lattice discretization. The $\Omega$
integration is compact and obeys the rules coming
from group theory. We have, for instance:
\be
\int {D\Omega\; \Omega _{ij}(x)\Omega
_{kl}^+}(y)=\delta _{xy}\delta _{il}\delta
_{jk}\label{quarantuno}
\ee where all the $\delta$'s are Kronecker
$\delta$'s, since we are on a lattice.

The correction linear in $W$, $\left\langle
{O_{GI}} \right\rangle _{(1)}$,  in
Eq.(\ref{expansion}) vanishes trivially in virtue
of the gauge average. On the contrary, for the
second order correction, 
$\left\langle {O_{GI}} \right\rangle _{(2)}$, we
have:
\bea
\left\langle {O_{GI}} \right\rangle _{(2)}\approx
a^2\int {dy_1dy_2\int {D\Omega \left\langle
{O_{GI}W^\Omega (y_1)W^\Omega (y_2)} \right\rangle
}}\approx\label{quarantadue}\nonumber\\
\approx a^{10}\sum\limits_{y_1,y_2} {\int {D\Omega
\left\langle {O_{GI}W^\Omega (y_1)W^\Omega (y_2)}
\right\rangle }}
\eea

In the last equality we resorted to the explicit
lattice notation for the integral in order to keep
track correctly of the powers of $a$.

From Eq.(\ref{quarantuno}) we know that the
$\Omega$-integration makes the two $W$  insertions
stick together, giving rise to a (non-)renormalized
gauge invariant operator  of dimension 10, $0_{10}$:
\bea
\left\langle {O_{GI}} \right\rangle _{(2)}\approx
a^{10}\sum\limits_{y_1} {\left\langle
{O_{GI}O_{10}(y_1)} \right\rangle }\approx
\nonumber\\
\approx a^6\int {dy_1\left\langle
{O_{GI}O_{10}(y_1)} \right\rangle
}\label{quarantatre}
\eea where we reintroduced the continuum notation.

The gauge-invariant operator $0_{10}$, defined by
Eq.(\ref{quarantatre}),
 mixes (with power divergent coefficients) with 
gauge-invariant operators of lower dimension:
\be O_{10}\approx \sum\nolimits_k {{1 \over
{a^{(10-k)}}}O_k}\label{quarantaquattro}
\ee

The factor $a^6$ in Eq.(\ref{quarantatre}) selects
from the mixing, defined
 in Eq.(\ref{quarantaquattro}), all the
gauge-invariant operators of dimension 4, $O_4$, or
less, with appropriate coefficients.

We thus get, for instance:
\be
\left\langle {O_{GI}} \right\rangle _{(2)}\approx
\int {dy_1\left\langle {O_{GI}O_4(y_1)}
\right\rangle }\label{average}
\ee
 
This procedure can be carried out order by order in
the $W$ insertion and the conclusion is  that the
effect of the gauge average in the computation of
gauge-invariant observables can be reabsorbed by a
redefinition of the parameters already present in
the gauge-invariant part of the action
Eq.(\ref{effective}), $S_{GI}$. This argument seems
to suggest that, after the gauge average, the
expectation value of any  gauge-invariant
observable, $O_{GI}$ will suffer again from the
doublers contribution.

Is this argument safe? This is not completely
clear. In fact, although this argument is
non-perturbative, the order by order expansion in
$W$ may be questioned. Certainly this  argument
could fail in presence of spontaneous symmetry
breaking. In fact in such  situations an explicit
breaking of the symmetry is needed to select a
particular vacuum  (the one aligned along the
direction of the breaking term) and the formal
expansion  in powers of the symmetry breaking term
could easily cause troubles connected  with the
failure of the cluster property.  In the gauge case
this should not cause any problem because we know,
from  Elitzur's theorem\cite{elitzur}
 that the local symmetry does not suffer
spontaneous symmetry breaking. 

It could, however, be argued that the expansion in
the Wilson term may  be non-analytic: after all the
Wilson term modifies in a dramatic way the physical
spectrum of the theory. Although this possibility
cannot be disproved  in general, it is instructive
to examine what happens in a very simple,
completely solvable case.

Consider a free fermion theory, in presence of the
Wilson term, written, for notational simplicity, in
the continuum language:
\be L=\int {dx\bar \psi (x)\not \partial \psi
(x)}+r a \int {dx\bar{\psi}_L\Box\psi_R + h.c.}
\ee

Suppose we want to compute, in such a theory, the
correlator $\Pi(q^2)$ of the vector current $j_\mu
(x)\equiv \bar \psi (x)\gamma _\mu \psi (x)$:
\be
\Pi(q^2)(q_\mu q_\nu -q^2\delta _{\mu \nu })
\equiv \int {d^4x \over (2\pi)^4}\left\langle
{j_\mu (x)j_\nu (0)} \right\rangle e^{iqx}
\ee

As well known $\Pi(q^2)$ has a logarithmic
divergence proportional to the number of particles
 running in the loop. Therefore the coefficient
will be different in the theory with $r\ne 0$ and
the one with $r=0$ because of the presence of the
doublers. If, when $r\ne 0$, we put:
\be
\Pi_{r\ne 0} (q^2)\approx \beta \log
(a^2q^2)+finite \; terms\label{quarantotto}
\ee then, in the case $r=0$ we have:
\be
\Pi_{r=0} (q^2)\approx 2^4\beta \log
(a^2q^2)+finite \; terms\label{quarantanove}
\ee precisely because in this case the doublers
will contribute.  In
Eqs.(\ref{quarantotto}),(\ref{quarantanove}), the
coefficient $\beta$ is independent of $r$, while
the {\em finite terms} in Eq.(\ref{quarantotto})
show a logarithmic singularity as $r \to
0$\cite{kawai}.

Let us see how this behaviour can be obtained by
expanding the  theory with $r \ne 0$ in powers of
$r$. We have:
\be
\Pi _{r\ne 0}(q^2)=\Pi
_{r=0}(q^2)+\sum\limits_{n=1}^\infty  {{r^n \over
{n!}}\Pi _{(n)}(q^2)}
\label{cinquanta}
\ee where $\Pi _{(n)}(q^2)$ denotes the
contribution to $\Pi (q^2)$ coming from the
insertion  of $n$ Wilson terms. These insertions
are infrared divergent\footnote{The argument is
completely analogous to the one used in the
computation of the effective
potential\cite{coleman}},  but the contribution to
the infrared divergence comes only by the doublers.
In fact the Wilson term is
 of order $q^2$ for $q^2\approx 0$, but is of order
$1$ for
$q^2$ around the momenta of any of the doublers. As
a consequence, for small $a$, we  have, (for $n >
0$):
\be {1 \over {n!}}\Pi _{(n)}(q^2)\mathop \approx
 \limits_{a\to 0}(-1)^{n+1}{{2^4-1} \over n}\beta
{1 \over {(a^2q^2)^n}}
\ee We get, therefore, from Eq.(\ref{cinquanta}):
\bea
\Pi _{r\ne 0}(q^2)\mathop \approx \limits_{a\to
0}\Pi _{r=0}(q^2)+ (2^4-1)\beta
\sum\limits_{n=1}^\infty  {{{(-1)^{n+1}} \over
n}{{r^n} \over {(a^2q^2)^n}}}\approx\nonumber\\
\approx 2^4\beta \log (a^2q^2)+(2^4-1)\beta \log
(1+{r \over
{(a^2q^2)}})\approx\label{cinquantadue}\\
\approx \beta \log (a^2q^2)\nonumber
\eea Eq.(\ref{cinquantadue}) shows that the
cancellation of the doubler contribution does not
necessarily require a non-analytic behavior in $r$.

\section{Fermion non conservation}

In the approach just oulined, only Fermion Number
conserving Green's functions  can be defined at the
Lattice level.

This does not necessarily imply Fermion
conservation.

In fact through Cluster Factorization we can define
Green's functions  related to Fermion violating
processes.

We may compute, for example:
\be
\left\langle {O_{\Delta F=2}(x)O_{\Delta F=-2}(y)}
\right\rangle
\ee and consider the limit:
\be
\mathop {\lim }\limits_{x-y\to \infty }\left\langle
{O_{\Delta F=2}(x)O_{\Delta F=-2}(y)} \right\rangle
=\left\langle {O_{\Delta F=2}(x)} \right\rangle
\left\langle {O_{\Delta F=-2}(y)} \right\rangle
\ee

It is, however, more consistent and interesting to
formulate\cite{redundant}  the theory from  the
start without redundant degrees of freedom,
corresponding to the  Target Theory:
\bea L=L_g+L_{gf}\nonumber\\ L_g=\bar \chi _{\dot
\alpha }\not D^{\dot \alpha \beta }\chi _\beta +{1
\over 4}F_{\mu \nu }^aF_{\mu \nu }^a\\ L_{gf}={1
\over {2\alpha _0}}\left( {\partial _\nu A_\nu ^a}
\right)\left( {\partial _\mu A_\mu ^a} \right)+\bar
c\partial _\mu D_\mu c\nonumber
\eea

The Fermionic euclidean functional integration is
now performed over the  independent Grassmann
variables $\bar \chi _{\dot \alpha }$ and $\chi
_\alpha $.

In the discretization process we have again the
doubling phenomenon, and it can  be avoided through
a Majorana-Wilson term of the form:
\be L_W=a\left( {\chi ^\alpha \partial _\mu
\partial _\mu \chi _\alpha +\bar \chi ^{\dot \alpha
}\partial _\mu \partial _\mu \bar \chi _{\dot
\alpha }} \right)
\ee

The Majorana-Wilson term is still ÒirrelevantÓ and
induces finite or power  divergent non-Lorentz
invariant and Fermion number violating counterterms
with $D\le 4$,  to be fixed again by BRST
identities.

This scheme has been checked in 1-loop perturbation
theory by G.Travaglini\cite{majorana}  and it works
fine.

Within such a formulation it is now possible to
write down fermion number violating  Green's
functions that are order $a$ in perturbation
theory, but can be enhanced and  promoted to finite
objects in presence of non-perturbative
configurations as, for example,  instantons.

\section {Concluding Remarks}

$\bullet$	Wilson-Yukawa theories were not discussed
in this talk, but they can be (and, in fact, have
been) treated along the same lines\cite{borrelli2}.

$\bullet$ Fine Tuning and Naturalness.

The problems outlined in this talk are not
necessarily merely technical. In fact if fine tuning
would turn out to be a really necessary ingredient
for the definition of a Chiral  Gauge Theory, this
may cast serious doubts on the Naturalness concept
in a purely field theoretical framework, with
possible far reaching implications  on the nature
of the more fundamental theory of which quantum
field theory could be considered
 a low energy approximation.

$\bullet$ Gribov problems?

Within the Rome approach to Chirality a rather
fundamental technical ingredient  is represented by
the gauge fixing. The presence of the ambiguities
due to the Gribov
 phenomenon still represents a serious challenge
within a non-perturbative framework. Certainly much
more (difficult) work has to be done in order to
clarify this important issue.

\section*{Acknowledgments}

I thank the organizers of the APCTP-ICTP Joint
International Conference  on Recent Developments in
Nonperturbative Quantum Field Theory for the kind
invitation.

A warm thank goes also to Professors Luciano Maiani
and Giancarlo Rossi for innumerable and very
fruitful discussions on the subject of the present
talk, over the past few years.

\end{document}